\documentclass[lettersize,journal]{IEEEtran}
\usepackage{amsmath,amsfonts}
\usepackage{amssymb}
\usepackage{mathtools}
\usepackage{xfrac}
\usepackage{algorithmic}
\usepackage{algorithm}
\usepackage{array}
\usepackage[caption=false,font=normalsize,labelfont=sf,textfont=sf]{subfig}
\usepackage{textcomp}
\usepackage{stfloats}
\usepackage{url}
\usepackage{verbatim}
\usepackage{enumitem}
\usepackage[breakwithin]{parnotes}
\makeatletter
\def\parnoteclear{%
	\gdef\PN@text{}%
}
\makeatother
\usepackage{tabularx}
\usepackage{graphicx}
\graphicspath{{img/}} % set graphics/image path
\usepackage{cite}
\usepackage{nomencl}
\begin{document}

\title{Reliability and Preventive Maintenance\\of Ducted Wind Turbines}

\author{Shafat Sharar, \textit{Clarkson University}\\
		Carl D. Hoover, \textit{Clarkson University}
        % <-this % stops a space
\thanks{S. Sharar is graduate student at Department of Mechanical and Aerospace Engineering, Clarkson University, New York, USA}% <-this % stops a space
\thanks{C. D. Hoover is faculty at Department of Mechanical and Aerospace Engineering, Clarkson University, New York, USA}}

% The paper headers
%\markboth{Journal of \LaTeX\ Class Files,~Vol.~14, No.~8, August~2021}%
%{Shell \MakeLowercase{\textit{et al.}}: A Sample Article Using IEEEtran.cls for IEEE %Journals}

%\IEEEpubid{0000--0000/00\$00.00~\copyright~2021 IEEE}
% Remember, if you use this you must call \IEEEpubidadjcol in the second
% column for its text to clear the IEEEpubid mark.

\maketitle

\begin{abstract}
This paper presents a reliability life analysis and preventive maintenance schedule for ducted wind turbines. Ducted wind turbines (DWT) are an emerging segment of the renewable energy industry with innovations that promise reliable, efficient, low-cost energy for consumer and small business markets. Many attempts have been made to build viable ducted turbines over the last century, but until recently none have succeeded commercially. Optimal shroud and blade designs are the focus of most engineering research to improve performance and efficiency, however, we hypothesize that an equally important key to the long-term success of small wind innovations is reliability analysis. For consumers and companies who want to efficiently maximize the lifespan of DWTs, this has significant ramifications. Operating beyond service life can result in catastrophic component failure and high replacement costs, making the technology economically infeasible. Our approach is focused on the analysis of 3.5 kW D3 turbines manufactured by Ducted Wind Turbines, Inc. We develop a component-level reliability analysis using ASTM E3159 and a consumer-level preventative maintenance schedule including failure modes and life estimates. Future research can use these findings to guide options for DWT life extension as well as localized maintenance solutions meant to reduce operational costs while preserving energy output.
\end{abstract}

\begin{IEEEkeywords}
reliability, preventive maintenance, maintenance schedule, ducted wind turbine
\end{IEEEkeywords}

\section{Introduction}

\IEEEPARstart{R}{enewable} Energy is a developing energy sector with continued global warming trends and energy demands driving research to make it more reliable and accessible. Recent trends show that renewable energy adoption in the forms of solar, hydropower, and small wind have increased since 2019. Total wind power capacity in the US has increased from approximately 105 GW to 135 GW in 2 years \cite{acp_clean_nodate} and is further projected to exceed 400 GW by 2050 \cite{noauthor_map_nodate}. According to the Global Wind Energy Commission (GWEC), the rate of wind energy adoption has increased in the last 6 years totaling a capacity of 743 GW worldwide \cite{noauthor_global_2021}. It is also expected that the wind energy market will have an average compound annual growth rate (CAGR) of 4\% each year. Different engineering optimizations of wind energy production are being researched to improve efficiency and make wind power more accessible to consumers. Ducted wind turbines (DWT) are one new technology that promises highly efficient energy generation in a compact form. \cite{noauthor_technology_nodate}

A `duct' on a turbine is an annular concentric shroud around the turbine that helps channel airflow to improve efficiency. The shroud cross-section typically has an airfoil-shaped profile. Ongoing studies at Clarkson University have focused on the aerodynamic performance of ducted turbines \cite{bagheri-sadeghi_maximal_2021, kummer_use_2020, valyou_design_2020}. Solar energy, while being popular and still having an upward trend, becomes ineffective during nighttime \cite{akker_case_2012}. Since energy storage is costly, a wind turbine in conjunction can offset solar inefficiency. The small wind sector has seen increased research and development investments, and with the continuing trend of shifting to 100\% renewable energy, small wind is projected to be an integral part of the common household in the US by 2050 \cite{noauthor_map_nodate}.

The IEC defines \textit{small wind} turbines as having a rotor-swept area of up to $200 m^2$ generating power at a voltage below 1000 V AC or 1500 V DC \cite{internationale_elektrotechnische_kommission_design_2019}. They typically produce between 500 W and 10 kW. However, power ranges vary depending on multiple factors and local standards. One of these factors is the presence of a duct around the rotor. The duct concentrates the flow enabling higher rotor rotation and efficient power generation compared to open turbines \cite{kanya_experimental_2018}.

Ducted wind turbines are more compact and are less susceptible to environmental damage, compared to open-rotor turbines. Recent developments and optimizations in DWTs have shown promising results. For example, the selection of the rotor power coefficient as a design parameter changes the nature of the annual energy production (AEP) optimization, which leads to turbine size minimization compared to open rotor turbines \cite{valyou_design_2020}. The key cost driver in this study was lowering turbine thrust, which both reduces the cost and increases AEP.

Further design optimization can be the type of airfoil used. An example study has been carried out by Kummer \cite{kummer_use_2020} where cambered plate airfoils were studied and compared to conventional and flat airfoils. The results showed that cambered plate airfoils can reduce the manufacturing cost greatly for a similarly performing conventional airfoil at Reynolds number of $40,000$.

Predictive Maintenance (PdM) typically refers to a data-driven tool that uses machine learning (ML) and statistical models to predict system failure trends and behavioral patterns to enable early fault detection and optimal maintenance plans, to avoid downtime \cite{sezer_industry_2018}. Preventive maintenance (PM) on the other hand is maintenance on a regular predetermined intervals to minimize the chance of component failure from regular use. It is useful to discuss the concept of \textit{life-limiting} which is to replace \textit{in-service} components after a predetermined life. This is usually done for components having increasing failure rates with age \cite{e11_committee_guide_nodate}. The maintenance schedule is often specified contractually. Usage outside the suggested scenario may shorten system life. Using a maintenance schedule geared towards commercial wind is not recommended for small wind systems as it may be costly and unrealistic for potential customers. An ML model aims to provide better insight into product behavior and information on making \textit{on-demand} maintenance schedules for DWTs.

A repairable system is defined as one that can be repaired and reused several times before retiring. On the contrary, a non-repairable system is retired after its first failure. A key metric for repairable systems is the mean time between failures (MTBF), the mean failure time between failures. Mean time to failure (MTTF) refers to the average life duration up to the first failure. MTBF applies to all sequences of repair and restoration cycles over a service life period. Another metric termed  $B_p$ life represents the p-th percentile of the life distribution. For example, $B_{10}$ life corresponds to a time value, $t$ when the cumulative distribution function (CDF), $F(t)$ is equal to $10\%$. In general, though no existing turbine specifications are presently available, a DWT aims to achieve a high MTBF and $B_p$ life.
\nomenclature{$B_p$ life}{$p$-th percentile of life distribution}
\nomenclature{MTTF}{Mean time to failure}
\nomenclature{MTBF}{Mean time between failures}

An appropriate measure of \textit{life} needs to be determined before assessing reliability. Operating time, cycles of usage, demand cycles, and calendar days are similarly valid measures of life but relate to different activities. For example, ‘cycles’ may last for any duration for however many cycles whereas ‘calendar time’ denotes continuous usage over that time. Life for less complex systems is generally expected to be longer, as fewer sub-systems and/or components are prone to failure. Small wind systems such as DWTs have fewer components compared to their larger, open-rotor counterparts. This leads to choosing the appropriate units for life measurements, whether years, weeks, days, or hours. On a component level, however, multiple lifing measures and/or units may be used for the sake of calculation. These different units can then be unified with recommended usage conditions. One point to note during calculations is that the duty cycle may vary depending on the system configuration of the DWT, e.g. braked or running, affecting its service life.

\begin{figure}[!t]
	\centering
	\includegraphics[width=2.5in]{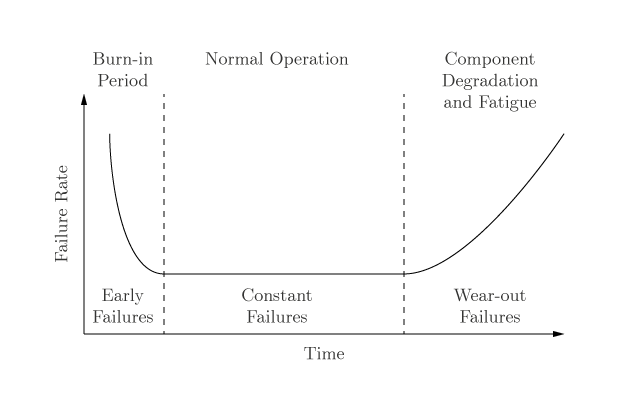}
	\caption{The Bathtub curve is a general representation of type and possibility of failures over time.}
	\label{fig:bathtub}
\end{figure}

Of the three general failure modes, infant mortality or early failures occurring in some units due to special causes can be avoided by conducting a 
``burn-in'' application during a ``break-in'' period. Random failures in wind turbines may be caused by collisions with birds, lightning strikes, flying debris during strong gusts or hurricanes, etc \cite{ma_wind_2018, noauthor_most_nodate, noauthor_metropolitan_nodate}. The blades are the most vulnerable to these external failures. For example, a total of 8 turbines had catastrophic failures in China because of typhoon Usagi in 2013 \cite{chen_structural_2016}. Random-type failure modes are assumed for reliability planning in the design phase. Wear-out failure mode is generally caused by gradual performance degradation. In wind turbines, the most likely worn parts are moving parts such as the yaw bearings, rotor bearings, and slip rings.  In modeling these failure modes for each component or subsystem, a broad approach is to make use of the ‘bathtub curve’ (shown in figure \ref{fig:bathtub}) which depicts the failure rate over time. For a wind turbine system, where the failure rate is variable, the average failure rate over an interval can be calculated. Wind speed distribution can be modeled by a Weibull curve \cite{ouarda_probability_2015, dhiman_chapter_2020} from which the instantaneous failure rate curve as well as the average failure rate over an interval can be calculated in closed form.

In determining reliability, a test plan is set up which generally consists of sample size, the test duration, and a life requirement. Reliability requirements are generally met at some confidence level such as 90\% or 95\%. In case the confidence is unspecified, it is assumed to be 50\%. Life requirements imply a measure of life and reliability at the stated life. For economical and timing purposes, an accelerated test is used where usage severity is higher than regular usage. This is done by tuning parameters and variables according to available accelerated models such as the Power-Weibull model \cite{nelson_accelerated_2004}. This model can be applied to ball or roller bearings testing, which are common components in wind turbines. A distribution assumption may be made based on the data properties, engineering knowledge, or prior performance. Furthermore, based on the parameters of the assumed distribution, additional variations will exist.

Field performance data are the principal indicators of reliability. Frequency, timing, severity, and cause of failure occurrences are the key reliability intelligence known as reliability data. Bench testing is done in development activity, but it may be difficult to emulate all possible field conditions, which is the case for DWTs. Often the case for such emerging technologies during bench or field testing is \textit{right suspension}. If a unit is functioning without failure after a specific time, it is called a right suspension or a ``run-out''. Hence the failure time which is now unknown, is in the future or to the right of the timeline.

The DWT system is a set of interconnected and interacting components and subsystems that function as a whole. Although systems can have multiple configurations, they have two fundamental types, series, and parallel. A series configuration is analogous to a chain made up of links. The chain system fails if at least one of the links fails. Whereas a parallel system will continue functioning if at least one of the components functions. Complex systems can combine multiple series and parallel configurations. In a simple analysis of system reliability, the individual component reliability is assumed to be independent of each other. So, it is possible to calculate the system reliability from the reliability of each component or subsystem. This simplification can be made to a certain degree based on the system properties. It is to be noted that, in calculating reliability, the failure probability may also be used as it is complementary to it.

There are several ways of calculating reliability or failure depending on the selected failure type, reliability measure, and available data. The probability plotting technique is the most appropriate for the DWT reliability data. The \textit{assumed} cumulative distribution function (CDF) versus the time plot is scaled so that a straight line can be approximated. The slope of the line corresponds to the respective parameter of the assumed distribution. The system MTTF is a function of the rate of each component. The easiest way to calculate system MTTF and associated distribution is to use Monte Carlo simulation.

The configuration of the blade pitch angle induced by the effect of rotor tilt affects blade fatigue life as determined by Hoghooghi \cite{hoghooghi_individual_2020} from testing a sub-scale model. This may be a useful lifing factor depending on the rotor configuration of current and future iterations of DWTs.

The remaining useful life (RUL) is the usable life left on a component at a specific time during operation. Depending on the type of repair, the RUL of a unit may either increase, decrease or remain the same. For example, a simple renewal process would restore the unit almost to its new condition. It is important to count the number of repairs during a service life period, which also includes replacements. Repair events are exponential for random type failures and the number of events is Poisson distributed.

The DWT is manufactured by Ducted Wind Turbines, Inc. and exclusively uses patented technology from Clarkson University proven as a commercially feasible small wind turbine. The purpose of the duct is to maximize energy extraction from the air that passes through the rotor. The efficiency metric is the power coefficient, $C_p$, which is the ratio of electrical power output to the wind power input. The Betz limit is known as the limiting factor to the efficiency in classical open rotor turbines, where maximum power coefficient, $C_{p,max}$ is $\sfrac{16}{27}$. The research from Jamieson \cite{jamieson_generalized_2008} shows that augmenting airflow with a duct affects this limit such that-

\begin{equation*}
	C_{p,max} = \frac{16}{27} (1 - a_0)
\end{equation*}

where, $a_0$ is the \textit{axial induction factor} (the fractional decrease in wind velocity between the free stream and the rotor plane) at the rotor plane. For open rotor systems, $a_0$ is $0$. The total power coefficient, however, remains within the Betz limit. A significant benefit of the duct is that it comparatively reduces the overall size of the turbine making it easier to install and maintain by customers. Masukume states that lack of commercialization and appropriate engineering techniques for maintenance are shortcomings of small ducted turbines \cite{masukume_technoeconomic_2014}. The engineering and optimizations in DWT design eliminates the economical drawbacks of historical small wind turbines by reducing cost per kilowatt electricity generated \cite{valyou_design_2020}.
\nomenclature{$C_{p}$}{Power coefficient}
\nomenclature{$C_{p, max}$}{Maximum power coefficient}
\nomenclature{$a_0$}{Axial induction factor}

\begin{figure}[!t]
\centering
\includegraphics[width=2.5in]{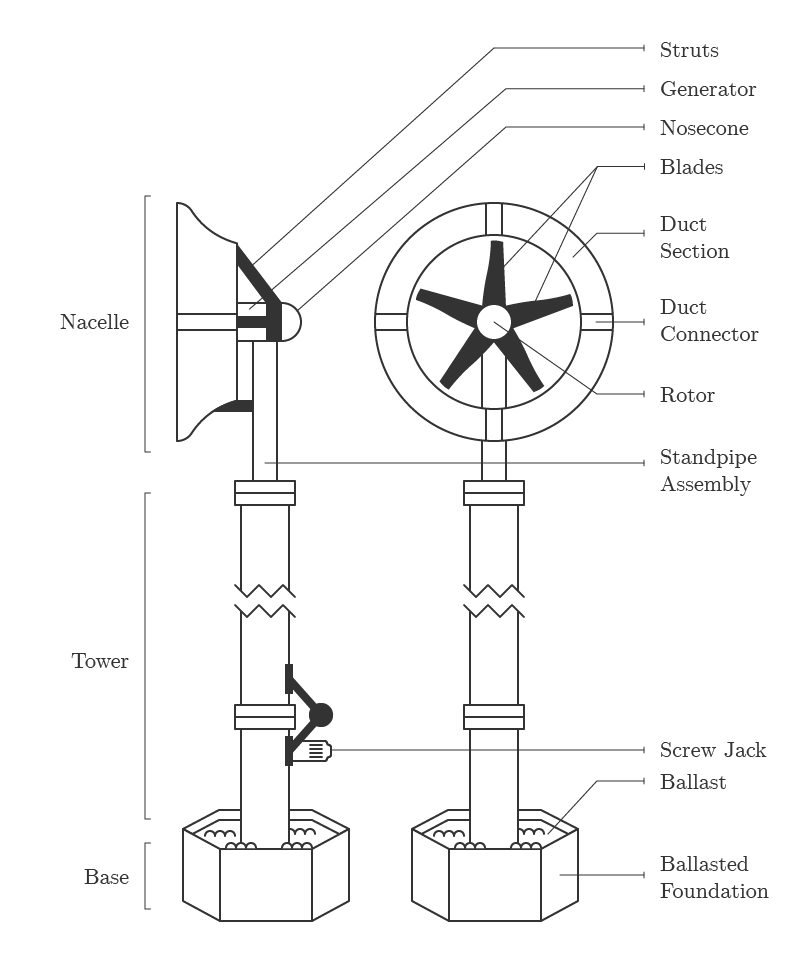}
\caption{Schematic diagram of the Ducted Wind Turbine with visible components labeled.}
\label{fig:dwt_sketch}
\end{figure}

The DWT can be segmented into 3 major sections, the foundation, the tower or pole, and the nacelle. Figure \ref{fig:dwt_sketch} shows a labeled diagram of the DWT. The foundation is the first step to installing a DWT, on which the tower stands. The base part of the tower is installed with the foundation, while the rest of it is assembled with the nacelle and finally lifted up with the help of a screw jack, which remains attached for easy lowering and/or raising of the turbine in the future. The nacelle and tower are coupled via a slewing bearing, which enables yaw action.

In the Ducted Wind Turbines, Inc. D3 ducted turbine, the shroud or duct is composed of four equal duct sections connected via custom-fabricated connectors.  The connectors are made of aluminum with female sockets on either side corresponding to recessed male ends on each duct sector. The male ends of each duct accept the connectors such that the surface of the duct remains as smooth as possible. 32 screws secure each connector to the duct sections via rivnuts installed in the duct. The duct is rotationally molded and made of proprietary PVC composite material. Its width is roughly 0.6 m and the enclosed diameter is roughly 3.7 m. The duct not only concentrates airflow but the airfoil structure of the cross-section also acts as a passive yaw controller aligning the rotor axis direction to that of the wind. In current designs, there are a total of 5 pressed aluminum blades connected to the rotor, which then couples directly to the generator. As blade designs evolve, blade materials and types are expected to change. A slip ring connects the wiring from the generator throughout the tower to the controller after passing through a shunt. The shunt is an electrical braking system that is employed if the turbine needs to be stopped for any reason. This braking is controlled by the controller unit.

\begin{figure}[!t]
	\centering
	\includegraphics[width=2.5in]{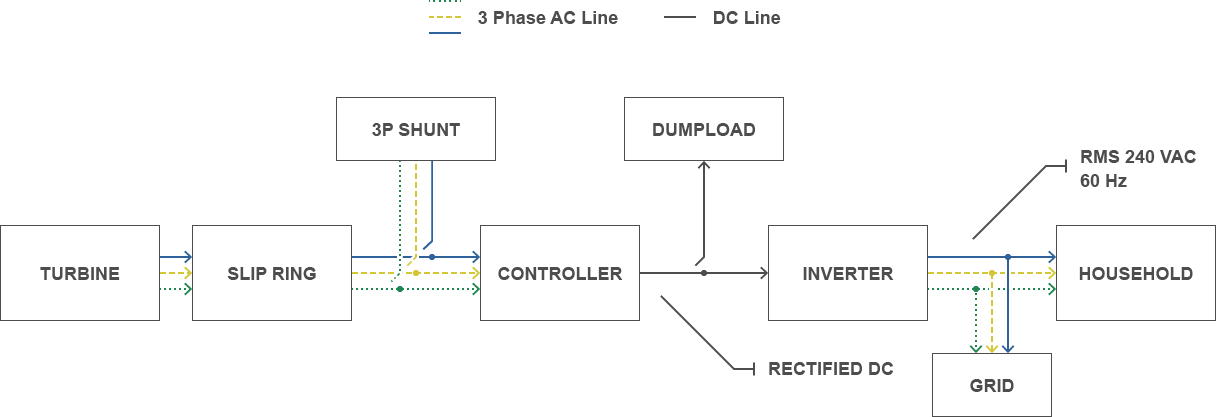}
	\caption{Schematic of the power flow in a typical DWT unit. The three lines represent a 3-phase AC line, while single line represents DC.}
	\label{fig:power_flow}
\end{figure}

Power transmission is done similarly to transmission in other renewable energy technologies. Generator power is transmitted to a controller which rectifies it and routes to either the brake resistors or the inverter. The inverter regulates this power for the consumer power line and the grid as needed. A simplified flow chart is shown in Figure \ref{fig:power_flow} for the power transmission in a typical DWT unit. A control system is built around the electromechanical system. This monitors the components and other sensors attached to the DWT, and controls the overall unit.

\section{Preventive Maintenance}
 
Preventive Maintenance (PM), or more commonly known as \textit{Scheduled Maintenance}, is a routine or scheduled inspection and maintenance put in place to identify developing or developed irregularities in system components. The goal of this practice is to repair these irregularities or replace the affected component(s) before they cause major failure to the system. The prime directive of PM is to improve system reliability and increase component life. Benefits of PM are system downtime reduction, system RUL enhancement, operators and maintenance mechanics' safety, and overall process cost reduction \cite{levitt_complete_2011}.

PM is an established practice employed in industrial systems as well as consumer equipment. The way PM works is by following a maintenance schedule. It documents a time-frame for the schedule, the required maintenance tasks, the typical operating conditions, and relevant comments to further improve the scheduling. The maintenance schedule is the result of an engineering analysis on reliability and life of the system components. The DWT system is a relatively simple one as discussed in the introduction. This part of the paper discusses reliability analysis on some of the major DWT components with a goal of building an initial maintenance schedule. The component analysis is anchored around the ASTM E3159-21, `Guide for General Reliability' \cite{e11_committee_guide_nodate}.

First, we identify and list the components of the DWT down to the fasteners. A bill of materials (BOM) from Ducted Wind, Inc. helps in this regard. The list is grouped by the four DWT subsystems `Structural', `Electromechanical', `Control', and `Fasteners'. While fasteners are integral to the other systems, we consider them as one distinct system because their analysis is fairly uniform. The analysis uses manufacturer specified life whenever available, documenting rated conditions. Some components, however, either have no manufacturer-recommended service data or are custom manufactured. This section of the paper analyzes those components and produces an in-depth analysis with relevant assumptions and operating conditions.

\subsection{Foundation}

The ballasted foundation contains sand and gravel which are subjected to constant weathering and erosion. This type of foundation requires replenishment of ballast when the level reaches a predetermined low. This weight is uniformly distributed on the base and provides foundational strength required to counteract the tipping moment of the tower. The ballast weight is found from an analysis by Daniel Valyou\footnote{Daniel Valyou is the Chief Engineer at Ducted Wind Turbines, Inc.} by taking the sum of moments on the tower.

\begin{flalign*}
	&\sum M = 0 \\
	\implies &T_{\text{turbine}} d_{\text{turbine}} - W d_{\text{base}} = 0 \\
	\implies &W = T_{\text{turbine}} d_{\text{turbine}} n / d_{\text{base}}
\end{flalign*}
\nomenclature{$W$}{Weight of nacelle}
\nomenclature{$T_{\text{turbine}}$}{Turbine thrust}
\nomenclature{$d_{\text{turbine}}$}{Largest diameter of nacelle}
\nomenclature{$n$}{Factor of Safety}
\nomenclature{$d_{\text{base}}$}{Foundation base diameter}

where, $T_{\text{turbine}}$ is the turbine thrust, $d_{\text{turbine}}$ is the largest diameter of the nacelle, $n$ is the factor of safety and $d_{\text{base}}$ is the foundation base diameter. The weight of the ballast is a function of the ballast volume, $V_{\text{ballast}}$, which in turn is a function of depth, $h_{\text{ballast}}$. So the weight is the product of density and volume-

\begin{align*}
	W &= \rho_{\text{ballast}} V_{\text{ballast}} = f(V_{\text{ballast}}) \tag*{[density of ballast is constant]}
\end{align*}
\nomenclature{$\rho_{\text{ballast}}$}{Ballast density}
\nomenclature{$V_{\text{ballast}}$}{Ballast volume}
\nomenclature{$h_{\text{ballast}}$}{Ballast height}

And the volume is the product of the cross-sectional area of the base and the depth-

\begin{align*}
V_{\text{ballast}} &= A_{\text{base}} h_{\text{ballast}} = f(h_{\text{ballast}}) \tag*{[area of the base is constant]} \\
\therefore W &= f(h_{\text{ballast}})
\end{align*}
\nomenclature{$A_{\text{base}}$}{Cross-sectional area of the base}

Hence, the depth or level can be a linear function of weight. The maintenance task then is to inspect the level of the ballast every year and replenish it as necessary to maintain the required weight. A reinforced cement concrete (RCC) foundation has a longer life and does not require replenishment. However, it is subject to erosion causing permanent damage. Especially when  weather conditions cause frequent freezing and thawing. The repair for the RCC foundation is to remove and recast.

\subsection{Tower/Pole}

The tower is connected to the foundation. This acts as a compression member with the vertical eccentric loading from the weight of the nacelle and the moment loading from the yaw rotation at the slewing bearing.

The analysis report by Valyou further evaluates a finite element analysis (FEA) model of a DWT unit at the Council Rock Enterprises, Inc. facility in Rochester, NY. The loads used are for a $109$ mph ($48.72 ms^{-1}$) 50-year extreme gust loading case with 3-second gusts. Valyou selects a worst-case loading direction for the tower where the moment arm is minimum from the tower center line and the perimeter. For the selected operating conditions the total turbine thrust is $6035$ N. With a hub height of $1.934$ m, the moment at the tower turbine mating flange is $11672$ N-m. The material ASTM A36 steel is uniform throughout the tower. The determined loads from the report are $1572$ lb-f load from the nacelle weight, $100$ lb-f load along the tie rods, and $81,500$ ft-lbs tipping moment about the axis perpendicular to azimuth and rotor axes. The FEA simulation results showed maximum values of 1st and 3rd principal stresses and Von Mises stress as $16.57$ ksi, $8.67$ ksi, and $\sigma' = 13.56$ ksi respectively.

With the analysis in place, we can verify the results using the formula for a column with eccentric loading.

\begin{gather}
	\frac{d^2 y}{dx^2} + \frac{P}{EI}y = - \frac{Pe}{EI}\\
	\delta = e \left[ \sec{\sqrt{\frac{P}{EI}} \frac{1}{2}} - 1\right]\\
	\frac{P}{A} = \frac{S_{yc}}{1 + \frac{ec}{k^2} \sec{\left(\frac{l}{2k} \sqrt{\frac{P}{AE}}\right)}}
\end{gather}
\nomenclature{$P$}{Load}
\nomenclature{$E$}{Young's modulus}
\nomenclature{$I$}{Moment of inertia}
\nomenclature{$e$}{Eccentricity}
\nomenclature{$\delta$}{Deflection}
\nomenclature{$c$}{Centroidal distance}
\nomenclature{$k$}{Radius of gyration}
\nomenclature{$l$}{Column height}
\nomenclature{$ec/k^2$}{Eccentricity ratio}

This is the \textit{secant column formula}. $P$ is the eccentric load, $E$ is Young's modulus for the tower material, $I$ is the moment of inertia, $\delta$ is column deflection, and $l$ is the height. Here $ec/k^2$ is known as the eccentricity ratio in which, $e$ is eccentricity, $c$ is centroidal distance in the tower cross section (typically a circle), and $k$ is the radius of gyration.

For the lifing estimate, we use the fatigue life method. The endurance limit is defined by-

\begin{gather}\label{eq:Se}
	S_e' = 0.5 S_{ut}\\
	S_e = k_a k_b k_c k_d k_e k_f S_e'
\end{gather}
\nomenclature{$S_e'$}{Endurance limit}
\nomenclature{$S_e$}{Endurance limit based on operating conditions}
\nomenclature{$S_{ut}$}{Minimum ultimate tensile strength}
\nomenclature{$k_a$}{Surface factor}
\nomenclature{$k_b$}{Size factor}
\nomenclature{$k_c$}{Loading factor}
\nomenclature{$k_d$}{Temperature factor}
\nomenclature{$k_e$}{Reliability factor}
\nomenclature{$k_f$}{Miscellaneous factor}

where $S_{ut}$ is the minimum ultimate tensile strength which is 58 ksi, $k_a$ through $k_f$ denote Marin modification factors related to the operating conditions under which the endurance limit is modified to be $S_e$. For the tower in axial loading and a 95\% reliability, the modification factors are shown in table \ref{tab:marin_mods}.

\begin{table}[H]
	\centering
	\caption{Marin Endurance Limit Modification Factors for Tower}
	\label{tab:marin_mods}
	\begin{tabular}{|c|l|c|}
		\hline
		\textbf{Factor} & \textbf{Condition} & \textbf{Value} \\
		\hline
		$k_a$ & Machined surface & 0.92 \\
		\hline
		$k_b$ & No size effect & 1.00 \\
		\hline
		$k_c$ & Axial loading & 0.85 \\
		\hline
		$k_d$ & Ambient temperature & 1.00 \\
		\hline
		$k_e$ & 95\% desired reliability & 0.87 \\
		\hline
		$k_f$ & No miscellaneous effects & 1.00 \\
		\hline
	\end{tabular}
\end{table}

Plugging these values result in an endurance limit of 19.68 ksi. We can use these values to estimate the remaining life with the following equations-

\begin{subequations}\label{eq:cycles}
\begin{gather}
	N = \left( \frac{\sigma'}{a} \right)^{1/b} \label{eq:N}\\
	a = \frac{(f S_{ut})^2}{S_e} \label{eq:a}\\
	b = -\frac{1}{3} \log \left( \frac{f S_{ut}}{S_e} \right) \label{eq:b}
\end{gather}
\end{subequations}
\nomenclature{$f$}{Fatigue strength fraction}

Here $f$ is the fatigue strength fraction which we take as 0.9. Solving for the constants $a$ and $b$ and plugging them in to $N$ results in approximately 14M cycles. This is true only in case of fully reversed stress, denoted by $\sigma'$. The stress from the FEA report is a combined stress so we expect the tower life to be much greater than this. We can define a cycle as complete rotations of the nacelle, or oscillations of average wind direction angle changes. If we assume the tower going through 1000 fully reversed stress cycles per day, this will translate to 365,000 cycles a year or a life of over 38 years.

\subsection{Pole Raising System}

The screw jack is a pole-raising and lowering system. The turbine can be lowered for securing during hazardous weather conditions or for maintenance and raised afterward. The manufacturer lists maintenance tasks of lubrication and dust filter replacement every 15 cycles, where a cycle is referred to as 1 raising and 1 lowering task combined. They estimate life at 100 cycles or 5 years, whichever comes first. After the manufacturer specified life, the jack is to be inspected for wear in gearing and screws. In the case of a custom-manufactured pole raising system where a maintenance schedule is unavailable, or if a sub-component is replaced, the life is a function of the life of the installed gears and screws.

The possible failure modes are determined from the specified maintenance tasks and sub-components. For example, a dust filter signals a possibility of accumulation of debris that can interfere with the lubrication and increase friction causing wear in the gearbox. Additionally, the motor operating the screw jack can fail independently, so, it is imperative to use a compatible screw jack and tower.\\

\subsection{Standpipe Assembly}

The standpipe assembly is a welded steel component that couples with the inner slewing bearing ring. The stress concentration points are near the flanges where the curvature is the lowest and are the main locations of wear and crack formation. The hub integrates into the top of this assembly and the generator sits inside the hub. A previous iteration of the DWT used aluminum as the material for the standpipe assembly. The component sagged in due to the load causing the attached strut to sag with it. This caused one of the blades to hit the duct causing the blade to bend and fail. This event shows the importance of material selection and tracking relative component position and resulting vibrations. An accelerometer may be fitted to these components for this purpose. We discuss more possible sensors for tracking maintenance and failures in the later chapters.

\subsection{Slewing Bearing}

The slewing bearing is the major component in the yaw motion of the DWT. It affects the direction of the rotor and the amount of energy extracted from the wind. Slewing bearings primarily carry axial and moment loads. The make and model of the bearing used in the current iteration of DWT is the \textit{Silverthin STO-145T}.

We follow a slewing bearing design guide for the equations. Moreover, we consult a recent study on fatigue failure of slewing bearings that works with a similar bearing to the one in the DWT. The National Renewable Energy Laboratory in the United States published their yaw and pitch rolling bearing design guideline \cite{harris_wind_2009} which specifies the basic dynamical axial load rating as follows-
\begin{equation}
C_a = f_{cm}(i \cos{\alpha_c})^{0.7} Z^{2/3} D_r^{1.8} \tan{\alpha_c}
\end{equation}
where $f_{cm}$ is the raceway groove factor which is the ratio of the groove radius and ball diameter, $D_r$. $Z$ is the number of balls, $i$ is the rolling element row, and $\alpha_c$ is the ball and raceway contact angle. However, since the bearing in DWT is not a continuously rotating type, but rather an oscillating type, the basic dynamic axial load is modified (as per regular operating conditions) as follows-
\begin{equation}
	C_{a,osc} = C_a \left(\frac{180^\circ}{\theta}\right)^{1/p}
\end{equation}
Here $\theta$ is half of the total arc traced during a cycle of oscillation and $p$ is 3 for ball bearings. We conservatively assume $\theta=30^\circ$ from the average turbine/wind direction data. It is important to note that this range of oscillations occurs during gusts, so life should be greater than the conservative result.

The dynamic equivalent axial load rating is shown below-
\begin{equation}
P_{ea} = 0.75 F_r + F_a + 2M / d_m
\end{equation}
Here, $F_r$ is radial load, $F_a$ is axial load, $M$ is the overturning moment, and $d_m$ is the bearing diameter up to the raceway center. Finally, the life can be estimated with the $L_{10}$ formula which is defined as-
\begin{equation}
L_{10} = \left(\frac{C_{a,osc}}{P_{ea}}\right)^3
\end{equation}
We correct for the life estimate using ANSI/ABMA standard modified rating life formula-
\begin{equation}
	L_{nm} = a_1 a_2 a_3 a_4 L_{10}
\end{equation}
Here, the lifing modification factors $a_1$ through $a_4$ are for reliability, bearing steel material, lubrication and flexible supporting structure respectively.
\nomenclature{$C_a$}{Dynamical axial loading}
\nomenclature{$f_{cm}$}{Bearing raceway groove factor}
\nomenclature{$D_r$}{Bearing roller/ball diameter}
\nomenclature{$d_m$}{Raceway center diameter}
\nomenclature{$Z$}{Number of balls in bearing}
\nomenclature{$\alpha_c$}{Contact angle}
\nomenclature{$P_{ea}$}{Dynamic equivalent axial loading}
\nomenclature{$F_r$}{Radial load}
\nomenclature{$F_a$}{Axial load}
\nomenclature{$M$}{Overturning moment}
\nomenclature{$L_{10}$}{$L_{10}$ life}
\nomenclature{$a_1$}{Reliability modification factor}
\nomenclature{$a_2$}{Material modification factor}
\nomenclature{$a_3$}{Lubrication modification factor}
\nomenclature{$a_4$}{Supporting structure modification factor}
\nomenclature{$L_{nm}$}{ANSI/AGMA modified rating life}

With most of the dimensional data found from available CAD models of the bearing, we perform the calculations required. Assumptions are that the raceway groove factor conforms to that of 0.53. Also that the bearing uses common materials. We assume a reliability of 90\%. Table \ref{tab:ansi_mods} shows the geometric data and modification factor values used. The results show that the bearings have a modified life of 1.49M oscillations. This translates to a theoretical number of raceway stress cycles of $T = L_{nm} Z = 44.89$M cycles. The bearing oscillates with wind direction which we estimate to be about 1,500 times per day on average. This translates to about 550,000 per year or a life of 80 years.

\begin{table}[!t]
	\centering
	\caption{ANSI/AGMA Life Modification Factors}
	\label{tab:ansi_mods}
	\begin{tabular}{|c|l|c|}
		\hline
		\textbf{Factor} & \textbf{Condition} & \textbf{Value} \\
		\hline
		$a_1$ & 90\% desired reliability & 1.00 \\
		\hline
		$a_2$ & Material hardness HRC 58 & 1.00 \\
		\hline
		$a_3$ & Recommended life modification factor & 0.10 \\
		\hline
		$a_4$ & Tubular tower & 0.85 \\
		\hline
	\end{tabular}
\end{table}

He et al. conducted a small sample test with slewing bearings and cross-examined the experimental, theoretical, and simulation-based results \cite{he_fatigue_2018}. The test was run with an axial force of 330 kN, an overturning moment of $1.38 \times 10^5$ kN-mm, and a test speed of 4 rpm, for 16 days. The theoretical calculation using the equations discussed above was within 5.18\% and the FE-SAFE simulation was within 10.17\% of the fatigue test results.

Among the findings from the study, it was noticed that the test platform produced an anomalous noise halfway through the testing and the accelerometer readings reflected a similar anomaly. This provides insight into the use of effective sensors on the DWT, such as accelerometers can be used similarly to how He used them in the test setup, or a sound sensor can be mounted to the tower which is analogous to the test frame where the abnormal sound occurred. The collected data would be useful for early failure prediction of the bearing, given a model is made from the test data.

\subsection{Blades}

The blades of the DWT are one of the most crucial components and are prone to failure, especially random type failures such as getting hit by flying debris. Historical failures of blades have been observed where one of the blades bent catastrophically from colliding with the duct due to blade misalignment. The blade is custom designed and manufactured which is assumed to be a rectangle for the sake of simplicity. Further assumptions are the blades are cantilever beams with its gravitational load due to weight, a torsion from the blade design, and a thrust load at the free end. We evaluate each load separately.

The blades can be approximated as having dimensions of $L = 1498.8 mm$, $b = 185 mm$, and $t = 4 mm$. They are attached to the rotor at an angle $\phi$ about the radius line. The material is Aluminum 6061 T6, with tensile strength, $S_{ut} = 310$ MPa or $45$ ksi, and a density of 2.7 g/cc. Finding the Marin modified endurance limit for this material gives $S_e = 17.6$ ksi. The constants $a$ and $b$ from equation \ref{eq:cycles} are 93.196 and -0.1206 respectively. The Marin life modification factors are shown in the Table \ref{tab:marin_mods2}.

\begin{table}[!t]
	\centering
	\caption{Marin Endurance Limit Modification Factors for Blade}
	\label{tab:marin_mods2}
	\begin{tabular}{|c|l|c|}
		\hline
		\textbf{Factor} & \textbf{Condition} & \textbf{Value} \\
		\hline
		$k_a$ & Pressed surface & 1.01 \\
		\hline
		$k_b$ & Beam in bending & 0.89 \\
		\hline
		$k_c$ & Bending loading & 1.00 \\
		\hline
		$k_d$ & Ambient temperature & 1.00 \\
		\hline
		$k_e$ & 95\% desired reliability & 0.87 \\
		\hline
		$k_f$ & No miscellaneous effects & 1.00 \\
		\hline
	\end{tabular}
\end{table}

The bending moment from weight of the beam and the resulting stress are calculated as follows-

\begin{gather}
	M_{blade} = -F_{blade} L\\
	\sigma_{blade} = \frac{yM_{blade}}{I}
\end{gather}

Here, $M_{blade}$ is the moment at the blade root, $F_{blade}$ is the load due to weight, $\sigma_{blade}$ is the bending stress, and $I$ is the moment of inertia for the assumed rectangular cross section. The weight of one blade is approximately 3 kg, which creates a bending moment of 22055 N-mm about the bending axis. The orientation of the cross section depends on the angle of the blade mounting. In the loading cases of a horizontal cross section, the bending stress is 44.7 MPa, whereas a vertical cross section has a bending stress of 0.97 MPa. With the rotation of the rotor, the maximum stress occurs when the blades are in a horizontal position where the bending effects are maximum. There is zero bending moment when the blades are vertical (upright or hanging straight down). So this bending stress is reversible. We take the worst case of 44.7 MPa or 6.48 ksi stress for the life calculation.

Now we solve for the equation \ref{eq:cycles} to get the number of cycles $N$. This results in $4\times10^{9}$ cycles. On average, the rotor rotates at 100 rpm during regular operation\footnote{as per data logged on April 12, 2022}. From the logged frequency data, the daily number of cycles is 144,000. Given one rotor rotation equates to one blade load cycle, the blade would have a life of over 75 years.

For the thrust loading, the maximum shear stress occurs at the center line. So we calculate the maximum shear from the following equation.

\begin{gather}
	\sigma_{s, max} = \frac{T}{\alpha b t^2}
\end{gather}

where $\sigma_{s, max}$ is the maximum shear stress, $T$ is the applied torque, $\alpha$ is an empirical constant. Alternatively, we use the following equation with an error of less than 4\%.
\nomenclature{$M_{blade}$}{Bending moment in blade}
\nomenclature{$F_{blade}$}{Load on blade}
\nomenclature{$\sigma_{blade}$}{Bending stress in blade}
\nomenclature{$L, b, t$}{Length, width, and thickness}
\nomenclature{$\sigma_{s, max}$}{Maximum shear stress}
\nomenclature{$T$}{Applied torque}

\begin{gather}\label{eq:shear}
\sigma_{s, max} = \frac{T}{b t^3} \left( 3 + 1.8 \frac{t}{b} \right)
\end{gather}

Valyou's analysis determines the rotor thrust to be approximately 2000 N for the 50-year extreme gust loading case with 3-second gusts. This load applies torque on the blades. We assume this load to be distributed to the five blades as 200 N on each blade. This translates to a torque, $T= 300$ N-m. The maximum shear stress then can be calculated to be 1.232 MPa or 0.1787 ksi. When the blades rotate past the standpipe wake region, they see this load as fluctuating and thus the torsional load may be considered fully reversible. Plugging the stress value in equation \ref{eq:cycles} yields a life of $3.4 \times 10^{22}$ cycles. A worst case twisting scenario is when the blades carry all the rotor thrust, which means 400 N on each blade, or a torque of 600 N-m. The shear stress is 0.0899 ksi. This results in a life of $1 \times 10^{20}$ cycles, which means that torsion does not have significant effect on blade fatigue.

In the case of the rotor thrust applies torque to a single blade, it will twist. We can calculate this torque by comparing the relations for power and torque coefficients. These equations are typically used for open rotor turbines, but the parameters used should work for a ducted turbine with manageable error. These relations are shown below.

\begin{equation}\label{eq:cp}
	C_p = 2 \frac{P_T}{\rho_{air} A_{\text{rotor}} V_{\infty}^3}
\end{equation}

\begin{equation}\label{eq:cq}
	C_q = 2 \frac{T}{\rho_{air} A_{\text{rotor}} V_{\infty}^2 R}
\end{equation}
\nomenclature{$C_q$}{Torque coefficient}
\nomenclature{$P_T$}{Power developed by turbine}
\nomenclature{$\rho_{air}$}{Air density}
\nomenclature{$A_{\text{rotor}}$}{Rotor cross section area}
\nomenclature{$V_{\infty}$}{Air stream velocity}
\nomenclature{$R$}{Rotor radius}

Where $C_p$ is the power coefficient of the turbine and $P_T$ is the power developed. $C_q$ is the torque coefficient and $T$ is the torque on the rotor. $R$ and $A_{\text{rotor}}$ are the rotor radius and cross-sectional area respectively. $V_{\infty}$ is the wind stream velocity and $\rho_{air}$ is the air density. We know that $P_T = T \omega$, where $\omega$ is the angular velocity of the rotor. Substituting this into the equations and dividing equation \ref{eq:cp} by equation \ref{eq:cq} yields the following.

\begin{equation}
	\frac{C_p}{C_q} = \frac{R \omega}{V}
\end{equation}

The wind speed at the free stream is $48.72 ms^{-1}$ and air density is $1.24 kg m^{-3}$. The DWT has a rotor diameter of 3 m. We assume the power coefficient and rotor speed for the conditions stated to be 0.40 and 600 rpm respectively. This results in a torque coefficient of 0.207. We can then calculate actual developed torque $T$, using equation \ref{eq:cq} which is approximately 3231 N-m. Using equation \ref{eq:shear} we solve for the shear stress on a single blade which is 13.269 MPa or 1.924 ksi. Again, due to similar reasons of the blades rotating past the standpipe's wake region, this load is reversible. Using the lifing equation \ref{eq:cycles} we get a cycle count of $2.7 \times 10^{15}$. This is still higher than what was for the bending moment. We can further vary our assumptions for power coefficient, $C_p$ and rotor speed, $N$, and estimate how the life changes based on that. Table \ref{tab:torque_life} shows the life estimates for different combinations of power coefficient and rotor speeds.

\begin{table}[!t]
	\centering
	\caption{Blade life variation due to varying turbine parameters.}
	\label{tab:torque_life}
	\begin{tabular}{|c|c|c|c|}
		\hline
		\textbf{$C_p$} & \textbf{$N$ (rpm)} & \textbf{$T$ (N-m)} & \textbf{Life (cycles)} \\
		\hline
		0.5 & 600 & 4027 & $1.5 \times 10^{13}$\\
		\hline
		0.6 & 600 & 4838 & $3.3 \times 10^{12}$\\
		\hline
		0.4 & 900 & 2154 & $2.7 \times 10^{15}$\\
		\hline
	\end{tabular}
\end{table}

As we can see from the life estimates in Table \ref{tab:torque_life}, the life is still long. The worst case loading then is due to bending moment from a horizontally mounted beam, meaning the twist angle is $\phi = 0^\circ$. This yields a life of about 75 years for the blades. Due to the complex geometry of the blade, this approximation introduces large errors. So we suggest using stresses from the FEA simulation to get a more accurate stress estimate, similar to the tower.

Despite the conservative analysis, it is recommended to conduct failure testing on the blades. One of the most common issues DWT faces is the blades not matching design and material tolerances. Accelerated testing of blades could provide invaluable data which can be used with a Weibull model to better determine reliability. For example, the current aluminum blades can be loaded and torqued with a higher load. Loads are scaled by a severity multiplication factor. The scaled variables are called accelerated variables. The time to failure is recorded for each blade. The accelerated variables will have a relationship to the typical usage time, which can be used to scale back the time during lifing estimation. The shape parameter ($\beta$) and scale parameter ($\eta$) are the basis to a Weibull distribution which can be defined by the cumulative distribution function (CDF)-

\begin{equation}\label{eq:weibull_cdf}
	F(t) = 1 - e^{-\left(\frac{t}{\eta}\right)^\beta}
\end{equation}
\nomenclature{$\beta$}{Weibull shape parameter}
\nomenclature{$\eta$}{Weibull scale parameter}

After testing and collecting results for two values of quantile life $B_p$ and $B_q$, equation \ref{eq:beta} is used to calculate the value of $\beta$.

\begin{equation}\label{eq:beta}
	\beta = \frac{\ln (-\ln (1 - p / 100)) - \ln (-\ln (1 - q / 100))}{\ln(B_p) - \ln(B_q)}
\end{equation}

The value of $\beta$ can then be used in equation \ref{eq:bp} to solve for $\eta$.

\begin{equation}\label{eq:bp}
	B_p = \eta \{ -\ln (1 - p / 100)\}^{1/\beta}
\end{equation}

With both the shape and scale parameters determined, we can create a Weibull probability plot which provides a substantial basis for determining reliability over time. A utility of Weibull distribution is shown in Predictive Maintenance chapter using wind speed. It is worth mentioning that in the design phase, a specific value of $\beta$ is set as a target to single out some design variables. Achieving that $beta$ value essentially means reaching the reliability goal.

\subsection{Nacelle}

The nacelle consists of the hub, the generator, the duct, the struts, the rotor, and the blades. The weight of the duct is distributed by struts that connect to the back of the hub and are supported by the standpipe, which in turn is supported by the slewing bearing. The struts act as the skeleton of the duct. Random-type failure is the most probable cause of failure for them. The 4 duct sections are connected via duct connectors which are aluminum sockets. The duct sections are rotational molded foam-filled polyvinyl chloride or PVC. Plastic materials thermally degrade when exposed to ultraviolet (UV) rays radiated from the sun. As a result, they become prone to damage from the elements causing failures such as cracking from hydrolysis. The same is true for the nose cone, which currently is a 3D-printed component made from acrylonitrile styrene acrylate or ASA.

\subsection{Generator}

The generator manufacturer Lancor documents the generator specifications on their brochure and website. The model is \textit{GSIP160}. The generator output is rated at 3500 W at the nominal shaft speed between 200 and 400 rpm. The insulation is rated as class F, which means that the maximum internal temperature is 130$^\circ$C. The lifing estimate provided by the manufacturer is 20 years at rated conditions. The generator is a direct drive meaning it is coupled directly to the rotor.

\subsection{Slip Ring}

The slip ring is manufactured by MOFLON, and its lifing and specifications are documented on the manufacturer’s website and product manual under the model number \textit{MW1630}. It is rated for 40M cycles at a rated speed of 250 rpm and 0.06 Nm torque. The ring oscillates with the change in yaw direction of the DWT. This implies that the number of revolutions is lower than for the rated life. So we expect the slip ring to last a greater number of cycles. Over current or over voltage can be a cause of burning out the slip ring as the wires are rated for a maximum of 600 V and 30 A.

\subsection{Fasteners}

The fasteners are assumed to last indefinitely as they were chosen with an appropriate safety factor in mind, and the only maintenance task is to retorque them periodically. Replacement would be necessary only if wear is visible during inspection. Since fasteners are non-reparable components, MTTF is the unit used for life.

\subsection{Controller}

The control system is composed of a controller, an inverter, a shunt brake, a dump load, and various communication cards used to extend the capabilities of the controller. The Voltsys controller controls the cut-in and cut-off speeds, applies brakes as necessary, logs sensory data, and channels power to the inverter and the dump load. The data contains sensory information from the generator, inverter, dump load, and anemometer. The controller manufacturer Voltsys lists the specifications for their base controller shown in Table \ref{tab:voltsys_specs}.

\begin{table}[!t]
	\centering
	\caption{Voltsys VS20A/800 Standard Unit Specifications}
	\begin{tabular}{|l|l|}
		\hline
		\textbf{Metric} & \textbf{Rated/Max Value} \\
		\hline
		Power Rating & 6 kW \\
		\hline
		Input AC Voltage & 600 V RMS 200 Hz \\
		\hline
		Input AC Current & 16.5 A\\
		\hline
		Output DC voltage & 800 V \\
		\hline
		Output DC current & 20 A\\
		\hline
		Diversion Load DC current & 20 A\\
		\hline
		Capacitance & 470 $\mu$F \\
		\hline
		AC Power Supply & 100-240 V (0.88 A) \\
		\hline
		Operating Temperature & -10$^\circ$C $\sim$ 40$^\circ$C \\
		\hline
	\end{tabular}
\label{tab:voltsys_specs}
\end{table}

We estimate the control system may fail the earliest mainly due to silicon/IC component degradation and technological obsolescence. As we discuss in later chapters, data has to be analyzed and accounted for as they are available. We assume the control systems will be adjusted and upgraded as per the software demand, which also requires that the assumption of components using silicon are actually user serviceable. Thus we can say that the control system life is limited by the life of the constituent silicon chips. Complementary metal–oxide–semiconductor or CMOS chip manufacturers in general use a design MTTF of 20 years \cite{noauthor_reliability_2000}. However, MTTF is not a strong indicator of reliability as it concerns the mean and not the failure rate. Pradeep \cite{ramachandran_metrics_2006} shows in his study of MTTF as a metric for CMOS reliability that a high percentage of chips fail before the designed MTTF of 20 years. As these chips are sourced from multiple manufacturers who uses varying level of quality control, a system can fail in as little as 10 years. The CDF plots from Pradeep's study support this conservative estimation for the controller life.

\subsection{System Life}

From the reliability analysis we estimate that the life of a DWT is limited by the generator, which is 20 years. We expect issues with the computing devices may arise the earliest, which is 10 years. Within assumed operating conditions we estimate the tower lasting about 38 years. The slewing bearing and the slip ring both will last about 80 years with recommended maintenance. The blades with nominal loading will last about 75 years. The non-repairable components have a service life of 1 year, meaning they have to inspected annually.
%\newpage
\section{Maintenance Schedule}

The study in the previous chapter concluded with determining RUL and key variables leading to maintenance tasks that can lead to prolonged life of DWT systems. This chapter in the study collects, tabulates and presents the aforementioned results, segmented by the 4 systems detailed earlier, as a maintenance schedule. The objective of this is to guide the DWT customers to a set of scheduled tasks helping maintain and prolong the RUL within recommended usage and working conditions.

The first 3 systems, i.e.,  structural, electromechanical, and control, contain diverse components which have specific failure modes and servicing tasks. These are, however, empirically established for the fasteners which are loaded within those standard conditions. So, the loaded fasteners are listed accompanied by a general maintenance schedule. The manufacturer recommended specifications are marked with asterisks (*).

% ---------------------------------
% BEGIN TABLE FOR STRUCTURAL SYSTEM
% ---------------------------------

\subsection{Structural System}

Table \ref{tab:structural} lists the components of the structural system of the DWT along with their respective failure modes and maintenance tasks. This system consists of everything that is supporting the nacelle and providing the structure required for the rotor to spin the generator.

\begin{table*}
\centering
\caption{Maintenance schedule for the `Structural' system of DWT\label{tab:structural}}
\begin{tabularx}{\textwidth}{
	| >{\hsize=0.5\hsize}X
	| >{\hsize=1\hsize}X
	| >{\hsize=0.5\hsize}X
	| >{\hsize=1.5\hsize}X
	| >{\hsize=1.5\hsize}X | }

%>{\hsize=.5\hsize}X

\hline
\parnoteclear
\textbf{Component} &
\textbf{Failure Modes} &
\textbf{Service Life} &
\textbf{Specifications} &
\textbf{Service Tasks} \\
\hline

Ballast Foundation\parnote{Ambor Structures: https://www.amborstructures.com/industries/engineered-structures} &
\begin{itemize}[leftmargin=*, nosep, before={\begin{minipage}[t]{\hsize}}, after ={\end{minipage}}]
\item Corrosion
\item Ballast loss
\item Loose bolts
\end{itemize} &
1 year &
\begin{itemize}[leftmargin=*, nosep, before={\begin{minipage}[t]{\hsize}}, after ={\end{minipage}}]
\item Maximum wind speed during maintenance: 38 mph*
\end{itemize} &
\begin{itemize}[leftmargin=*, nosep, before={\begin{minipage}[t]{\hsize}}, after ={\end{minipage}}]
\item Inspect ballast and hardware connections annually or upon high load event
\item Inspect bolt torques and service according to manual\parnote{Recommended Torque: https://www.amborstructures.com/wp-content/uploads/2018/08/QuikBase-Assembly-Manual.pdf}
\item Top off ballast material every year
\end{itemize} \\
\hline

Tower &
\begin{itemize}[leftmargin=*, nosep, before={\begin{minipage}[t]{\hsize}}, after ={\end{minipage}}]
	\item Bending from moment
	\item Fatigue from nacelle imbalance
\end{itemize} &
5 years* &
\begin{itemize}[leftmargin=*, nosep, before={\begin{minipage}[t]{\hsize}}, after ={\end{minipage}}]
	\item Max base moment: 55000 ft-lbs*
	\item Survival wind speed: 90 mph (3 sec gust)*
	\item Material: ASTM A572 GR65*
\end{itemize} &
\begin{itemize}[leftmargin=*, nosep, before={\begin{minipage}[t]{\hsize}}, after ={\end{minipage}}]
	\item Inspect column deflection, cracks, localized damage, every 5 years
	\item Inspect bolt torques and service according to manual\parnote{ARE Tower Manual: https://aretelecom.com/wp-content/uploads/2021/04/AFS-350-60ft-Monopole.pdf}
\end{itemize} \\
\hline

Screw Jack (Pole Raising System)\parnote{Manufacturer's Service Manuual: https://www.amborstructures.com/wp-content/uploads/2018/08/Ambor-Motorized-Assembly-Manual.pdf} &
\begin{itemize}[leftmargin=*, nosep, before={\begin{minipage}[t]{\hsize}}, after ={\end{minipage}}]
	\item Motor failure from weather exposure
	\item Gearbox damage/wear/corrosion
	\item Pin bending/corrosion
\end{itemize} &
5 years* &
\begin{itemize}[leftmargin=*, nosep, before={\begin{minipage}[t]{\hsize}}, after ={\end{minipage}}]
	\item Max wind speed (during maintenance): 38 mph*
\end{itemize} &
\begin{itemize}[leftmargin=*, nosep, before={\begin{minipage}[t]{\hsize}}, after ={\end{minipage}}]
	\item Inspect every 100 cycles or 5 years (whichever comes first)
	\item Inspect monthly for visible corrosion
	\item Lubricate every 15 cycles or once a year with recommended grease
	\item Replace dust cover
\end{itemize} \\
\hline

Standpipe Assembly &
\begin{itemize}[leftmargin=*, nosep, before={\begin{minipage}[t]{\hsize}}, after ={\end{minipage}}]
	\item Bolt failure
	\item Shearing
	\item Corrosion from gasket leakage
\end{itemize} &
10 years &
N/A &
\begin{itemize}[leftmargin=*, nosep, before={\begin{minipage}[t]{\hsize}}, after ={\end{minipage}}]
	\item Check flange bolt torque
	\item Replace gasket if corroded/worn
\end{itemize} \\
\hline

Hub &
\begin{itemize}[leftmargin=*, nosep, before={\begin{minipage}[t]{\hsize}}, after ={\end{minipage}}]
	\item Crack formation
	\item Chipping due to heat and vibration
\end{itemize} &
10 years &
N/A &
\begin{itemize}[leftmargin=*, nosep, before={\begin{minipage}[t]{\hsize}}, after ={\end{minipage}}]
	\item Inspect cracks/wear/pitting
\end{itemize} \\
\hline

Hub Bushing &
\begin{itemize}[leftmargin=*, nosep, before={\begin{minipage}[t]{\hsize}}, after ={\end{minipage}}]
	\item Crack formation
	\item Pitting
\end{itemize} &
10 years &
N/A &
\begin{itemize}[leftmargin=*, nosep, before={\begin{minipage}[t]{\hsize}}, after ={\end{minipage}}]
	\item Inspect cracks/wear/pitting
\end{itemize} \\
\hline

Blades &
\begin{itemize}[leftmargin=*, nosep, before={\begin{minipage}[t]{\hsize}}, after ={\end{minipage}}]
	\item Bending
	\item Shearing
	\item Impact from flying debris
\end{itemize} &
75 years &
\begin{itemize}[leftmargin=*, nosep, before={\begin{minipage}[t]{\hsize}}, after ={\end{minipage}}]
	\item Max operating wind speed: 130 mph
\end{itemize} &
\begin{itemize}[leftmargin=*, nosep, before={\begin{minipage}[t]{\hsize}}, after ={\end{minipage}}]
	\item Inspect cracks/wear/pitting
	\item Check alignment
\end{itemize} \\
\hline

Nose Cone &
\begin{itemize}[leftmargin=*, nosep, before={\begin{minipage}[t]{\hsize}}, after ={\end{minipage}}]
	\item Cracking
	\item Plastic Deformation
\end{itemize} &
10 years &
N/A &
\begin{itemize}[leftmargin=*, nosep, before={\begin{minipage}[t]{\hsize}}, after ={\end{minipage}}]
	\item Inspect cracks/wear/pitting
\end{itemize} \\
\hline

Struts (3 for duct support) &
\begin{itemize}[leftmargin=*, nosep, before={\begin{minipage}[t]{\hsize}}, after ={\end{minipage}}]
	\item Bending from compression
	\item Shearing at joints
\end{itemize} &
10 years &
N/A &
\begin{itemize}[leftmargin=*, nosep, before={\begin{minipage}[t]{\hsize}}, after ={\end{minipage}}]
	\item Inspect bending/deflection
\end{itemize} \\
\hline

Duct sockets &
\begin{itemize}[leftmargin=*, nosep, before={\begin{minipage}[t]{\hsize}}, after ={\end{minipage}}]
	\item Distortion
	\item Corrosion at plastic-metal interface
\end{itemize} &
10 years &
N/A &
\begin{itemize}[leftmargin=*, nosep, before={\begin{minipage}[t]{\hsize}}, after ={\end{minipage}}]
	\item Inspect corrosion/wear
	\item Clean rusting and use protectant (as required)
\end{itemize} \\
\hline

Duct sections &
\begin{itemize}[leftmargin=*, nosep, before={\begin{minipage}[t]{\hsize}}, after ={\end{minipage}}]
	\item Thermal degradation
	\item Cracking from hydrolysis
\end{itemize} &
20 years &
N/A &
\begin{itemize}[leftmargin=*, nosep, before={\begin{minipage}[t]{\hsize}}, after ={\end{minipage}}]
	\item Inspect cracks/wear
	\item Use UV protectant (as required)
\end{itemize} \\
\hline

Rain shielding &
\begin{itemize}[leftmargin=*, nosep, before={\begin{minipage}[t]{\hsize}}, after ={\end{minipage}}]
	\item Rusting
	\item Corrosion
\end{itemize} &
10 years &
N/A &
\begin{itemize}[leftmargin=*, nosep, before={\begin{minipage}[t]{\hsize}}, after ={\end{minipage}}]
	\item Inspect cracks/wear/pitting
\end{itemize} \\
\hline

\end{tabularx}
\begin{flushleft}
	\parnotes
\end{flushleft}
\end{table*}

% ----------------------------------------
% BEGIN TABLE FOR ELECTROMECHANICAL SYSTEM
% ----------------------------------------

\subsection{Electromechanical System}

The heart of the DWT is its electromechanical system. It consists of the generator, slew bearing, slip ring, and their related machine elements and wirings. The generator comes as a closed 3500 watt unit from the manufacturer and only has replaceable filters. This makes it compact but less repairable. It is rated at 200-400 rpm for power generation. The slew bearing and the slip ring are subjected to the most usage cycles within this system mechanically, and so are expected to fail earlier and more often relative to the other components. Table \ref{tab:electromechanical} shows the failure modes and maintenance tasks of the included components in this system.

\begin{table*}
\centering
\caption{Maintenance schedule for the `Electromechanical' system of DWT}
\label{tab:electromechanical}
\begin{tabularx}{\linewidth}{
	| >{\hsize=0.5\hsize\linewidth=\hsize}X
	| >{\hsize=1\hsize\linewidth=\hsize}X
	| >{\hsize=0.5\hsize\linewidth=\hsize}X
	| >{\hsize=1.5\hsize\linewidth=\hsize}X
	| >{\hsize=1.5\hsize\linewidth=\hsize}X | }

\hline
\parnoteclear
\textbf{Component} &
\textbf{Failure Modes} &
\textbf{Service Life} &
\textbf{Specifications} &
\textbf{Service Tasks} \\
\hline

Slew Bearing\parnote{Silverthin STO-145T: https://www.silverthin.com/bearings/slewing-rings/sto/detail/STO-145T} &
\begin{itemize}[leftmargin=*, nosep, before={\begin{minipage}[t]{\hsize}}, after ={\end{minipage}}]
	\item Raceway wear/pitting
	\item Surface spalling
	\item Thermal fatigue
\end{itemize} &
1 year &
\begin{itemize}[leftmargin=*, nosep, before={\begin{minipage}[t]{\hsize}}, after ={\end{minipage}}]
	\item Moment rating: 19050 ft-lbs
\end{itemize} &
\begin{itemize}[leftmargin=*, nosep, before={\begin{minipage}[t]{\hsize}}, after ={\end{minipage}}]
	\item Lubricate annually
	\item Replace (if necessary)
\end{itemize} \\
\hline

Slip Ring\parnote{MOFLON MW1630 manual: https://www.moflon.com/pdf/mw1630.pdf} &
\begin{itemize}[leftmargin=*, nosep, before={\begin{minipage}[t]{\hsize}}, after ={\end{minipage}}]
	\item Raceway wear/pitting
	\item Abrasion wear
	\item Thermal fatigue
\end{itemize} &
40 years &
\begin{itemize}[leftmargin=*, nosep, before={\begin{minipage}[t]{\hsize}}, after ={\end{minipage}}]
	\item Max speed: 250 rpm
	\item Torque: 0.06 Nm
	\item Operating Temp: $-40^\circ$C - $80^\circ$C
\end{itemize} &
\begin{itemize}[leftmargin=*, nosep, before={\begin{minipage}[t]{\hsize}}, after ={\end{minipage}}]
	\item Replace after 20M cycles
\end{itemize} \\
\hline

Slip Ring Wiring\parnote{16G cables} &
\begin{itemize}[leftmargin=*, nosep, before={\begin{minipage}[t]{\hsize}}, after ={\end{minipage}}]
	\item Burn-out
	\item Tearing of wire sleeve
	\item Shorting
	\item Splice failure
\end{itemize} &
N/A &
\begin{itemize}[leftmargin=*, nosep, before={\begin{minipage}[t]{\hsize}}, after ={\end{minipage}}]
	\item Max voltage: 600 VDC/VAC
	\item Max current: 30 A/wire
	\item Electrical noise: 10 m$\Omega$ @10 rpm
\end{itemize} &
\begin{itemize}[leftmargin=*, nosep, before={\begin{minipage}[t]{\hsize}}, after ={\end{minipage}}]
	\item Inspect wear
	\item Check terminal voltage
	\item Replace (if required)
\end{itemize} \\
\hline

Slip Ring collar &
\begin{itemize}[leftmargin=*, nosep, before={\begin{minipage}[t]{\hsize}}, after ={\end{minipage}}]
	\item Rusting/corrosion from leakage
\end{itemize} &
N/A &
N/A &
\par\noindent
\begin{itemize}[leftmargin=*, nosep, before={\begin{minipage}[t]{\hsize}}, after ={\end{minipage}}]
	\item Inspect annually
\end{itemize} \\
\hline

Generator Wiring to Slip Ring\parnote{Southwire SOOW 600V: https://www.southwire.com/wire-cable/flexible-cord/600v-royal-soow-cord-with-black-jacket-90-c/p/55809702} &
\begin{itemize}[leftmargin=*, nosep, before={\begin{minipage}[t]{\hsize}}, after ={\end{minipage}}]
	\item Burn-out
	\item Tearing of wire sleeve
	\item Shorting
	\item Splice failure
\end{itemize} &
N/A &
\begin{itemize}[leftmargin=*, nosep, before={\begin{minipage}[t]{\hsize}}, after ={\end{minipage}}]
	\item Max Amperage: 30 A
\end{itemize} &
\begin{itemize}[leftmargin=*, nosep, before={\begin{minipage}[t]{\hsize}}, after ={\end{minipage}}]
	\item Inspect Annually
\end{itemize} \\
\hline

Generator\parnote{Lancor 3kW Generator: https://www.lancor.es/wp-content/uploads/2021/02/Generator-brochure-GSIP-160-280-350-650-libro.pdf} &
\begin{itemize}[leftmargin=*, nosep, before={\begin{minipage}[t]{\hsize}}, after ={\end{minipage}}]
	\item Transmission shaft failure
	\item Thermal fatigue
	\item Demagnetization
	\item Filter wear
	\item Stator core damage
\end{itemize} &
20 years &
\begin{itemize}[leftmargin=*, nosep, before={\begin{minipage}[t]{\hsize}}, after ={\end{minipage}}]
	\item Rated output: 3 kW
	\item Nominal speed: 200 - 400 rpm
	\item Insulation class: F
\end{itemize} &
\begin{itemize}[leftmargin=*, nosep, before={\begin{minipage}[t]{\hsize}}, after ={\end{minipage}}]
	\item Inspect filters annually
	\item Replace filter (if necessary)
	\item Assess performance after 20 years
\end{itemize} \\
\hline

Generator Shaft &
\begin{itemize}[leftmargin=*, nosep, before={\begin{minipage}[t]{\hsize}}, after ={\end{minipage}}]
	\item Bending
	\item Vibration from keyway wear
	\item Misalignment from vibration
\end{itemize} &
20 years &
N/A &
\begin{itemize}[leftmargin=*, nosep, before={\begin{minipage}[t]{\hsize}}, after ={\end{minipage}}]
	\item Lubricate
	\item Clean rust
	\item Use protectant (as required)
\end{itemize} \\
\hline

PVC Conduit &
\begin{itemize}[leftmargin=*, nosep, before={\begin{minipage}[t]{\hsize}}, after ={\end{minipage}}]
	\item Wear
	\item Cracking
\end{itemize} &
1 year &
$\sfrac{3}{4}-6''$ &
\begin{itemize}[leftmargin=*, nosep, before={\begin{minipage}[t]{\hsize}}, after ={\end{minipage}}]
	\item Inspect annually
	\item Replace (if necessary)
\end{itemize} \\
\hline

\end{tabularx}
\begin{flushleft}
	\parnotes
\end{flushleft}
\end{table*}

% ------------------------------
% BEGIN TABLE FOR CONTROL SYSTEM
% ------------------------------

\subsection{Control System}

The control system consists primarily of the Voltsys controller which is the brain of the system coordinating the inverter and its modules, shunt brake, and dump loading. The components in this system are mainly electronics and thus their failure criteria are nearly identical. Excess generated power is channeled to the dump load or resistors and as a result they heat up significantly depending on the power. That is why they are more prone to thermal failure compared to the other components in the system. There can be multiple dump resistors, but the other components are single quantities. The following table contains the maintenance table for the control system.

\begin{table*}
\centering
\caption{Maintenance schedule for the `Control' system of DWT} 
\label{tab:control}
\begin{tabularx}{\linewidth} {
	| >{\hsize=0.5\hsize\linewidth=\hsize}X
	| >{\hsize=1\hsize\linewidth=\hsize}X
	| >{\hsize=0.5\hsize\linewidth=\hsize}X
	| >{\hsize=1.5\hsize\linewidth=\hsize}X
	| >{\hsize=1.5\hsize\linewidth=\hsize}X | }

\hline
\parnoteclear
\textbf{Component} &
\textbf{Failure Modes} &
\textbf{Service Life} &
\textbf{Specifications} &
\textbf{Service Tasks} \\
\hline

Voltsys Controller\parnote{Voltsys: https://www.voltsys.com/voltsys-20a-wind-hydro-controller/} &
\begin{itemize}[leftmargin=*, nosep, before={\begin{minipage}[t]{\hsize}}, after ={\end{minipage}}]
	\item Capacitor failure
	\item Integrated circuit (IC) failure
\end{itemize} &
N/A &
N/A &
\begin{itemize}[leftmargin=*, nosep, before={\begin{minipage}[t]{\hsize}}, after ={\end{minipage}}]
	\item Inspect after high load events
\end{itemize} \\
\hline

Dump Resistors &
\begin{itemize}[leftmargin=*, nosep, before={\begin{minipage}[t]{\hsize}}, after ={\end{minipage}}]
	\item Overheating/Burnout
\end{itemize} &
N/A &
N/A &
\begin{itemize}[leftmargin=*, nosep, before={\begin{minipage}[t]{\hsize}}, after ={\end{minipage}}]
	\item Inspect after high load events
\end{itemize} \\
\hline

Fimer Inverter\parnote{Fimer UNO-DM-6.0-TLPLUS-Q: https://www.fimer.com/system/files/2022-04/FIMER\_Brochure-Building\%20Applications-EN-revC.pdf} &
\begin{itemize}[leftmargin=*, nosep, before={\begin{minipage}[t]{\hsize}}, after ={\end{minipage}}]
	\item Overvoltage
	\item Overcurrent
	\item Capacitor failure
\end{itemize} &
5 years &
\begin{itemize}[leftmargin=*, nosep, before={\begin{minipage}[t]{\hsize}}, after ={\end{minipage}}]
	\item Output: 6000 W
	\item Max DC input power: 3500 W
	\item Max DC input current: 31.5 A
	\item Max external AC overcurrent protection: 40 A
	\item Operating temp: $-25^\circ$C - $60^\circ$C with derating above $45^\circ$C
\end{itemize} &
\begin{itemize}[leftmargin=*, nosep, before={\begin{minipage}[t]{\hsize}}, after ={\end{minipage}}]
	\item Inspect after high load events
	\item Inspect annually after 5 years
\end{itemize} \\
\hline

Comm Card\parnote{Communications card for Fimer inverters manufactured by `Inverter Supply'} &
\begin{itemize}[leftmargin=*, nosep, before={\begin{minipage}[t]{\hsize}}, after ={\end{minipage}}]
	\item IC failure
	\item Capacitor failure
\end{itemize} &
2 years &
\begin{itemize}[leftmargin=*, nosep, before={\begin{minipage}[t]{\hsize}}, after ={\end{minipage}}]
	\item Operating temp: $-20^\circ$C - $60^\circ$C
\end{itemize} &
\begin{itemize}[leftmargin=*, nosep, before={\begin{minipage}[t]{\hsize}}, after ={\end{minipage}}]
	\item Inspect after high load events
\end{itemize} \\
\hline

Shunt Brake &
\begin{itemize}[leftmargin=*, nosep, before={\begin{minipage}[t]{\hsize}}, after ={\end{minipage}}]
	\item Burnout
	\item Corrosion
\end{itemize} &
N/A &
N/A &
\begin{itemize}[leftmargin=*, nosep, before={\begin{minipage}[t]{\hsize}}, after ={\end{minipage}}]
	\item Inspect after high load events
\end{itemize} \\
\hline

\end{tabularx}
\begin{flushleft}
	\parnotes
\end{flushleft}
\end{table*}

% -------------------------
% BEGIN TABLE FOR FASTENERS
% -------------------------

\subsection{Fasteners}
The DWT contains various fasteners for different components and purposes. The bill of materials (BOM) for the QuickBase foundation provides the number of fasteners used including 6 anchor bolts with 4 nuts and washers to each for the center pedestal, and a total of 84 bolts with 1 nut and 2 washers to each for the bottom and side panels. The tower or pole mates with the pedestal with the anchor bolts. The parts of the tower are connected at the flanges via 8 bolts with 1 nut and 2 washers each. The standpipe assembly and the tower are connected via the slew bearing. This whole assembly requires a total of 32 bolts for the inner and outer rings of the bearing. The struts supporting the duct use 8 machine screws each. 16 rivnuts are inserted into the duct sections, which are used to mate the sections to the duct connectors with 16 machine screws each. The generator is connected to the hub of the standpipe assembly with a grub screw. The list of fasteners is in Table \ref{tab:fasteners} and their general maintenance schedule is in Table \ref{tab:fasteners_schedule}.

\begin{table*}
\centering
\caption{List of loaded fasteners used in DWT}
\label{tab:fasteners}
\begin{tabularx}{\linewidth} {
	| >{\hsize=0.75\hsize\linewidth=\hsize}X
	| >{\hsize=0.5\hsize\linewidth=\hsize}X
	| >{\hsize=0.75\hsize\linewidth=\hsize}X
	| >{\hsize=2\hsize\linewidth=\hsize}X | }

\hline
\textbf{Type/Size} &
\textbf{Grade/Class} &
\textbf{Specifications} &
\textbf{Details} \\
\hline

Bolt / M$12\times35$ &
Grade 8.8 &
Torque: 48 lb-ft* &
Ballast foundation bolts \\
\hline

Bolt / M$12\times35$ &
Grade 8.8 &
Torque: 119 lb-ft* &
Ballast foundation bolts \\
\hline

Bolt / M$33\times200$ &
Grade 8.8 &
Torque: 1083 lb-ft* &
Ballast foundation bolts \\
\hline

Bolt / M$36-4\times1000$ &
Grade 8.8 &
Torque: 777 lb-ft* &
Ballast foundation anchor bolts \\
\hline

Bolt / M$16-2\times60$ &
Grade 8.8 &
Torque: 119 lb-ft* &
Ballast foundation flange bolts \\
\hline

Bolt / M$16\times60$ &
Grade 8.8 &
Torque: 119 lb-ft* &
Screw jack bolts \\
\hline

Bolt / M$10\times35$ &
N/A &
Torque: 27 lb-ft* &
Screw jack bolts \\
\hline

Bolt / M$6\times50$ &
N/A &
N/A &
Screw jack bolts \\
\hline

Bolt / $\sfrac{5}{8}-11\times2\sfrac{1}{2}$'' &
Zn-Al Coated &
N/A &
Mfr: ARMOR COAT / Part No: UST235844 \newline
Connect tower to slew bearing and slew bearing to standpipe assembly \\
\hline

SHCS / M$10-1.5\times25$ &
Class 12.9 &
N/A &
Mfr: ARMOR COAT / Part No: UST710025 \newline
Generator attachment screws \\
\hline

Bolt / M$14-2\times50$ &
Class 8.8 &
N/A &
Mfr: Infasco / Part No: 67084 \newline
Generator hub bolt \\
\hline

SHCS / $\sfrac{3}{8}-16\times1\sfrac{1}{4}$'' &
Zn Plated &
N/A &
Mfr: ARMOR COAT / Part No: UST235957 \newline
Connect the blades to the hub \\
\hline

BHCS / \#$8-32\times\sfrac{1}{2}$'' &
Black Oxide &
N/A &
Nose cone screws \\
\hline

SHCS / $\sfrac{1}{4}-20\times\sfrac{5}{8}$'' &
Zn-Al Coated &
N/A &
Mfr: ARMOR COAT / Part No: UST235919 \newline
Connect struts to duct sockets and bedplate \\
\hline

SHCS / $\sfrac{1}{4}-20\times\sfrac{7}{8}$'' &
Zn-Al Coated &
N/A &
Mfr: ARMOR COAT / Part No: UST235921 \newline
Connect struts to duct sockets and bedplate \\
\hline

FBHCS / $\sfrac{1}{4}-20\times1$'' &
Zn Coated &
N/A &
Mfr: Socket Source / Part No: FBHA04C016US \newline
Connect bottom socket to duct(s) \\
\hline

Torx T25 Truss Head Machine Screws / $\sfrac{1}{4}-20$'' &
Zn Plated &
N/A &
Mfr: Jet Fitting / Part No: 1412MTT-2500 \newline
Connect duct sockets to ducts \\
\hline

Rivnut Plusnuts / $\sfrac{1}{4}-20$'' &
Yellow Zn Plated &
N/A &
Mfr: Libberty Engineering \newline
Seated into duct sections to mate with machine screws \\
\hline

Grub Screw / $\sfrac{3}{8}-16\times\sfrac{3}{4}$'' &
N/A &
N/A &
Mfr: McMasterr-Carr / Part No: N/A \\
\hline

Slip Ring Screw / \#$10-24\times\sfrac{1}{2}$'' &
N/A &
N/A &
Mfr: ARMOR COAT / Part No: UST235905 \newline
\\
\hline

SHCS / $\sfrac{1}{4}-20\times\sfrac{5}{8}$'' &
N/A &
N/A &
Mfr: ARMOR COAT / Part No: UST235919 \newline
Nacelle cover plate screws \\
\hline

Generator Shaft Key and Set Bolt &
N/A &
N/A &
Holds hub to generator shaft \\
\hline

\end{tabularx}
\end{table*}

\begin{table*}
\centering
\caption{Maintenance Schedule for fasteners in DWT}
\label{tab:fasteners_schedule}
\begin{tabularx}{\linewidth} {
	| >{\hsize=0.5\hsize\linewidth=\hsize}X
	| >{\hsize=1.25\hsize\linewidth=\hsize}X
	| >{\hsize=1.25\hsize\linewidth=\hsize}X | }

\hline
\textbf{Fastener Type} &
\textbf{Failure Modes} &
\textbf{Service Tasks} \\
\hline

Bolts (and Nuts) &
\begin{itemize}[leftmargin=*, nosep, before={\begin{minipage}[t]{\hsize}}, after ={\end{minipage}}]
	\item Loose bolts
	\item Shearing on nuts
	\item Bearing on bolts
	\item Rust/Corrosion
\end{itemize} &
\begin{itemize}[leftmargin=*, nosep, before={\begin{minipage}[t]{\hsize}}, after ={\end{minipage}}]
	\item Inspect
	\item Check torque
	\item Apply anti-rusting agent (as required)
	\item Replace (if necessary)
\end{itemize} \\
\hline

Screws &
\begin{itemize}[leftmargin=*, nosep, before={\begin{minipage}[t]{\hsize}}, after ={\end{minipage}}]
	\item Loose screws
	\item Shear
	\item Rust/Corrosion
\end{itemize} &
\begin{itemize}[leftmargin=*, nosep, before={\begin{minipage}[t]{\hsize}}, after ={\end{minipage}}]
	\item Inspect
	\item Apply anti-rusting agent (as required)
	\item Replace (if necessary)
\end{itemize} \\
\hline

\end{tabularx}
\end{table*}
\newpage
\section{Conclusion}
%1. Restate the thesis / main objective
DWT is an emerging technology that aims to achieve the longest life possible with minimum maintenance. This paper tried to touch on that goal by evaluating RUL and exploring reliability improvements of DWTs using both traditional and modern methods. Lifing analyses were conducted using ASTM standards and empirical machine design methods such as $B_p$ life, $L_{10}$ life, Weibull fitting, etc. This allowed us to build a simple maintenance schedule that is meant as a conservative way of maintaining the turbine with regular inspections.
%3. Explain why your work is relevant
The impact of this study is on the reliability aspect of ducted wind turbines. We conducted a component level life analysis which allowed for understanding of the variables affecting RUL. The biggest contribution is that a customer transferable maintenance schedule that had not existed before for the DWT.

One of the largest limitations this study has faced is the lack of historical failure data which left our proposed model unverified for DWT specific data. Also, being in the design phase, the DWT sees constant change in design and material selection which may nullify some of the component analysis. We tried to account for that using multiple methods of analysis for some of the major components. The equation parameters can easily be recalculated for any dimensional or material changes in the future.

\section{Future Work}
%4. A take-home message for the reader
This study on reliability just scratches the surface of a reliability growth study. We suggest doing failure testing for custom manufactured components which can be fit with the Weibull distribution. We expect more volume and variety of data to be available with such testing. This can be used to develop more suitable maintenance schedules. Furthermore, failure prediction models can be developed and validated which will bolster the reliability of DWTs.

%\begin{figure*}[!t]
%\centering
%\subfloat[]{\includegraphics[width=2.5in]{fig1}%
%\label{fig_first_case}}
%\hfil
%\subfloat[]{\includegraphics[width=2.5in]{fig1}%
%\label{fig_second_case}}
%\caption{Dae. Ad quatur autat ut porepel itemoles dolor autem fuga. Bus quia con nessunti as remo di quatus non perum que nimus. (a) Case I. (b) Case II.}
%\label{fig_sim}
%\end{figure*}

%{\appendices
%\section*{Proof of the First Zonklar Equation}
%Appendix one text goes here.
% You can choose not to have a title for an appendix if you want by leaving the argument blank
%\section*{Proof of the Second Zonklar Equation}
%Appendix two text goes here.}
\newpage
\bibliography{reference}

% Generated by IEEEtran.bst, version: 1.14 (2015/08/26)
\begin{thebibliography}{10}
\providecommand{\url}[1]{#1}
\csname url@samestyle\endcsname
\providecommand{\newblock}{\relax}
\providecommand{\bibinfo}[2]{#2}
\providecommand{\BIBentrySTDinterwordspacing}{\spaceskip=0pt\relax}
\providecommand{\BIBentryALTinterwordstretchfactor}{4}
\providecommand{\BIBentryALTinterwordspacing}{\spaceskip=\fontdimen2\font plus
\BIBentryALTinterwordstretchfactor\fontdimen3\font minus
  \fontdimen4\font\relax}
\providecommand{\BIBforeignlanguage}[2]{{%
\expandafter\ifx\csname l@#1\endcsname\relax
\typeout{** WARNING: IEEEtran.bst: No hyphenation pattern has been}%
\typeout{** loaded for the language `#1'. Using the pattern for}%
\typeout{** the default language instead.}%
\else
\language=\csname l@#1\endcsname
\fi
#2}}
\providecommand{\BIBdecl}{\relax}
\BIBdecl

\bibitem{acp_clean_nodate}
\BIBentryALTinterwordspacing
ACP, ``\BIBforeignlanguage{en-US}{Clean {Power} {Annual} {Market} {Report}
  2021}.'' [Online]. Available:
  \url{https://cleanpower.org/market-report-2021/}
\BIBentrySTDinterwordspacing

\bibitem{noauthor_map_nodate}
\BIBentryALTinterwordspacing
``\BIBforeignlanguage{en}{Map: {Projected} {Growth} of the {Wind} {Industry}
  {From} {Now} {Until} 2050}.'' [Online]. Available:
  \url{https://www.energy.gov/maps/map-projected-growth-wind-industry-now-until-2050}
\BIBentrySTDinterwordspacing

\bibitem{noauthor_global_2021}
\BIBentryALTinterwordspacing
``\BIBforeignlanguage{en-US}{Global {Wind} {Report} 2021},'' Mar. 2021.
  [Online]. Available: \url{https://gwec.net/global-wind-report-2021/}
\BIBentrySTDinterwordspacing

\bibitem{noauthor_technology_nodate}
\BIBentryALTinterwordspacing
``Technology {\textbar} {Ducted} {Wind} {Turbines} - {Changing} the {Face} of
  {Small} {Wind}.'' [Online]. Available:
  \url{https://www.ductedwind.com/technology}
\BIBentrySTDinterwordspacing

\bibitem{bagheri-sadeghi_maximal_2021}
\BIBentryALTinterwordspacing
N.~Bagheri-Sadeghi, B.~T. Helenbrook, and K.~D. Visser,
  ``\BIBforeignlanguage{en}{Maximal power per device area of a ducted
  turbine},'' \emph{\BIBforeignlanguage{en}{Wind Energy Science}}, vol.~6,
  no.~4, pp. 1031--1041, Jul. 2021. [Online]. Available:
  \url{https://wes.copernicus.org/articles/6/1031/2021/}
\BIBentrySTDinterwordspacing

\bibitem{kummer_use_2020}
\BIBentryALTinterwordspacing
A.~Kummer, J.~Dimeo, M.~Hebel, and K.~Visser, ``\BIBforeignlanguage{en}{On the
  {Use} of {Cambered} {Plate} {Airfoils} for {Small} {Wind} {Turbines}},''
  \emph{\BIBforeignlanguage{en}{Journal of Physics: Conference Series}}, vol.
  1618, no.~4, p. 042001, Sep. 2020. [Online]. Available:
  \url{https://iopscience.iop.org/article/10.1088/1742-6596/1618/4/042001}
\BIBentrySTDinterwordspacing

\bibitem{valyou_design_2020}
\BIBentryALTinterwordspacing
D.~N. Valyou and K.~D. Visser, ``\BIBforeignlanguage{en}{Design considerations
  for a small ducted wind turbine},'' \emph{\BIBforeignlanguage{en}{Journal of
  Physics: Conference Series}}, vol. 1452, no.~1, p. 012019, Jan. 2020.
  [Online]. Available:
  \url{https://iopscience.iop.org/article/10.1088/1742-6596/1452/1/012019}
\BIBentrySTDinterwordspacing

\bibitem{akker_case_2012}
J.~Akker, H.~Blok, C.~Budd, R.~Eggermont, A.~Guterman, D.~Lahaye,
  J.~Lansink~Rotgerink, K.~Myerscough, C.~Prins, T.~Tromper, and W.~Wadman, ``A
  {Case} {Study} in the {Future} {Challenges} in {Electricity} {Grid}
  {Infrastructure},'' Feb. 2012.

\bibitem{internationale_elektrotechnische_kommission_design_2019}
I.~E. Kommission, Ed., \emph{\BIBforeignlanguage{eng}{Design requirements}},
  edition 4.0~ed., ser. Wind energy generation systems / {International}
  {Electrotechnical} {Commission}.\hskip 1em plus 0.5em minus 0.4em\relax
  Geneva, Switzerland: International Electrotechnical Commission, 2019, no.
  part 1.

\bibitem{kanya_experimental_2018}
\BIBentryALTinterwordspacing
B.~Kanya and K.~D. Visser, ``\BIBforeignlanguage{en}{Experimental validation of
  a ducted wind turbine design strategy},'' \emph{\BIBforeignlanguage{en}{Wind
  Energy Science}}, vol.~3, no.~2, pp. 919--928, Dec. 2018. [Online].
  Available: \url{https://wes.copernicus.org/articles/3/919/2018/}
\BIBentrySTDinterwordspacing

\bibitem{sezer_industry_2018}
E.~Sezer, D.~Romero, F.~Guedea, M.~Macchi, and C.~Emmanouilidis, ``An
  {Industry} 4.0-{Enabled} {Low} {Cost} {Predictive} {Maintenance} {Approach}
  for {SMEs},'' in \emph{2018 {IEEE} {International} {Conference} on
  {Engineering}, {Technology} and {Innovation} ({ICE}/{ITMC})}, Jun. 2018, pp.
  1--8.

\bibitem{e11_committee_guide_nodate}
\BIBentryALTinterwordspacing
{E11 Committee}, ``\BIBforeignlanguage{en}{Guide for {General}
  {Reliability}},'' ASTM International, Tech. Rep. [Online]. Available:
  \url{http://www.astm.org/cgi-bin/resolver.cgi?E3159-21}
\BIBentrySTDinterwordspacing

\bibitem{ma_wind_2018}
Y.~Ma, P.~Martinez-Vazquez, and C.~Baniotopoulos, ``Wind {Turbine} {Tower}
  {Collapse} {Cases}: {A} {Historical} {Overview},'' \emph{ICE Proceedings
  Structures and Buildings}, vol. 172, May 2018.

\bibitem{noauthor_most_nodate}
\BIBentryALTinterwordspacing
``\BIBforeignlanguage{en}{Most common reasons for wind turbine failures}.''
  [Online]. Available:
  \url{https://www.cotes.com/blog/most-common-reasons-for-wind-turbine-failures}
\BIBentrySTDinterwordspacing

\bibitem{noauthor_metropolitan_nodate}
\BIBentryALTinterwordspacing
``\BIBforeignlanguage{en}{Metropolitan {Engineering} {Consulting} and
  {Forensics} - {CAUSE} {AND} {CONTRIBUTING} {FACTORS} {OF} {FAILURE} {OF}
  {GEARED} {WIND} {TURBINES}}.'' [Online]. Available:
  \url{https://sites.google.com/site/metropolitanforensics/cause-and-contributing-factors-of-failure-of-wind-turbines}
\BIBentrySTDinterwordspacing

\bibitem{chen_structural_2016}
\BIBentryALTinterwordspacing
X.~Chen and J.~Z. Xu, ``\BIBforeignlanguage{en}{Structural failure analysis of
  wind turbines impacted by super typhoon {Usagi}},''
  \emph{\BIBforeignlanguage{en}{Engineering Failure Analysis}}, vol.~60, pp.
  391--404, Feb. 2016. [Online]. Available:
  \url{https://linkinghub.elsevier.com/retrieve/pii/S1350630715301552}
\BIBentrySTDinterwordspacing

\bibitem{ouarda_probability_2015}
\BIBentryALTinterwordspacing
T.~Ouarda, C.~Charron, J.-Y. Shin, P.~Marpu, A.~Al-Mandoos, M.~Al-Tamimi,
  H.~Ghedira, and T.~Al~Hosary, ``\BIBforeignlanguage{en}{Probability
  distributions of wind speed in the {UAE}},''
  \emph{\BIBforeignlanguage{en}{Energy Conversion and Management}}, vol.~93,
  pp. 414--434, Mar. 2015. [Online]. Available:
  \url{https://linkinghub.elsevier.com/retrieve/pii/S0196890415000400}
\BIBentrySTDinterwordspacing

\bibitem{dhiman_chapter_2020}
\BIBentryALTinterwordspacing
H.~S. Dhiman, D.~Deb, and V.~E. Balas, ``\BIBforeignlanguage{en}{Chapter 2 -
  {Wind} energy fundamentals},'' in \emph{\BIBforeignlanguage{en}{Supervised
  {Machine} {Learning} in {Wind} {Forecasting} and {Ramp} {Event}
  {Prediction}}}, ser. Wind {Energy} {Engineering}, H.~S. Dhiman, D.~Deb, and
  V.~E. Balas, Eds.\hskip 1em plus 0.5em minus 0.4em\relax Academic Press, Jan.
  2020, pp. 9--21. [Online]. Available:
  \url{https://www.sciencedirect.com/science/article/pii/B9780128213537000132}
\BIBentrySTDinterwordspacing

\bibitem{nelson_accelerated_2004}
W.~Nelson, \emph{\BIBforeignlanguage{eng}{Accelerated testing: statistical
  models, test plans and data analysis}}.\hskip 1em plus 0.5em minus
  0.4em\relax Hoboken, N.J: Wiley, 2004.

\bibitem{hoghooghi_individual_2020}
\BIBentryALTinterwordspacing
H.~Hoghooghi, N.~Chokani, and R.~S. Abhari,
  ``\BIBforeignlanguage{en}{Individual {Blade} {Pitch} {Control} for {Extended}
  {Fatigue} {Lifetime} of {Multi}-{Megawatt} {Wind} {Turbines}},''
  \emph{\BIBforeignlanguage{en}{Journal of Physics: Conference Series}}, vol.
  1618, no.~2, p. 022008, Sep. 2020. [Online]. Available:
  \url{https://iopscience.iop.org/article/10.1088/1742-6596/1618/2/022008}
\BIBentrySTDinterwordspacing

\bibitem{jamieson_generalized_2008}
\BIBentryALTinterwordspacing
P.~Jamieson, ``\BIBforeignlanguage{en}{Generalized limits for energy extraction
  in a linear constant velocity flow field},''
  \emph{\BIBforeignlanguage{en}{Wind Energy}}, vol.~11, no.~5, pp. 445--457,
  2008, \_eprint: https://onlinelibrary.wiley.com/doi/pdf/10.1002/we.268.
  [Online]. Available:
  \url{https://onlinelibrary.wiley.com/doi/abs/10.1002/we.268}
\BIBentrySTDinterwordspacing

\bibitem{masukume_technoeconomic_2014}
\BIBentryALTinterwordspacing
P.-M. Masukume, G.~Makaka, and D.~Tinarwo,
  ``\BIBforeignlanguage{en}{Technoeconomic {Analysis} of {Ducted} {Wind}
  {Turbines} and {Their} {Slow} {Acceptance} on the {Market}},''
  \emph{\BIBforeignlanguage{en}{Journal of Renewable Energy}}, vol. 2014, p.
  e951379, Dec. 2014, publisher: Hindawi. [Online]. Available:
  \url{https://www.hindawi.com/journals/jre/2014/951379/}
\BIBentrySTDinterwordspacing

\bibitem{levitt_complete_2011}
J.~Levitt, \emph{Complete {Guide} to {Preventive} and {Predictive}
  {Maintenance}}, second edition~ed.\hskip 1em plus 0.5em minus 0.4em\relax New
  York, NY: Industrial Press Inc, 2011.

\bibitem{harris_wind_2009}
\BIBentryALTinterwordspacing
T.~Harris, J.~H. Rumbarger, and C.~P. Butterfield,
  ``\BIBforeignlanguage{en}{Wind {Turbine} {Design} {Guideline} {DG03}: {Yaw}
  and {Pitch} {Rolling} {Bearing} {Life}},'' Tech. Rep. NREL/TP-500-42362,
  969722, Dec. 2009. [Online]. Available:
  \url{http://www.osti.gov/servlets/purl/969722-YFwQR5/}
\BIBentrySTDinterwordspacing

\bibitem{he_fatigue_2018}
\BIBentryALTinterwordspacing
P.~He, R.~Hong, H.~Wang, and C.~Lu, ``\BIBforeignlanguage{en}{Fatigue life
  analysis of slewing bearings in wind turbines},''
  \emph{\BIBforeignlanguage{en}{International Journal of Fatigue}}, vol. 111,
  pp. 233--242, Jun. 2018. [Online]. Available:
  \url{https://linkinghub.elsevier.com/retrieve/pii/S0142112318300720}
\BIBentrySTDinterwordspacing

\bibitem{noauthor_reliability_2000}
\BIBentryALTinterwordspacing
``Reliability in cmos ic design : Physical failure mechanisms and their
  modeling,'' 2000. [Online]. Available:
  \url{https://www.mosis.com/files/faqs/tech_cmos_rel.pdf}
\BIBentrySTDinterwordspacing

\bibitem{ramachandran_metrics_2006}
\BIBentryALTinterwordspacing
P.~Ramachandran, S.~V. Adve, P.~Bose, J.~A. Rivers, and J.~Srinivasan,
  ``Metrics for {Lifetime} {Reliability},'' Aug. 2006. [Online]. Available:
  \url{https://hdl.handle.net/2142/11244}
\BIBentrySTDinterwordspacing

\end{thebibliography}
\bibliographystyle{IEEEtran}

%\newpage
%
%\section{Biography Section}
%
%\begin{IEEEbiographynophoto}{Shafat Sharar}
%is a graduate student in the Department of Mechanical and Aerospace Engineering at Clarkson University.
%\end{IEEEbiographynophoto}

\vfill

\end{document}